\documentclass[10pt,twocolumn,floatfix,altaffilletter,superscriptaddress,  tightenlines,showpacs,showkeys,preprintnumbers,nofootinbib]{revtex4-2}
\pdfoutput=1
\usepackage[colorlinks=true,citecolor=blue,linkcolor=blue,breaklinks=true]{hyperref}
\usepackage{amsmath,amssymb}
\usepackage{epsfig} 
\usepackage{graphicx}        
\usepackage{url}
\usepackage{color}
\usepackage{multirow}
\usepackage{placeins}
\usepackage[dvipsnames]{xcolor}
\usepackage{braket}
\usepackage{float}
\usepackage{slashed}
\usepackage{chngcntr}
\counterwithout{equation}{section}
\hypersetup{
  colorlinks=true,
  linkcolor=red,
  filecolor=magenta,   
  urlcolor=blue,
  citecolor=blue,
}

\allowdisplaybreaks

\bibpunct{[}{]}{,}{n}{}{,}

\def\beq{\begin{equation}}
\def\eeq{\end{equation}}
\def\bea{\begin{eqnarray}}
\def\eea{\end{eqnarray}}
\makeatletter
\begin{document}

%\title{Probing GeV scale right-handed neutrinos with inflationary gravitational waves and pulsar timing data}
\title{Fingerprint of GeV scale right-handed neutrinos on inflationary gravitational waves and PTA data }
 \author{Satyabrata Datta}
 \email{satyabrata.datta@saha.ac.in}
 \affiliation{Saha Institute of Nuclear Physics, 1/AF, Bidhannagar, Kolkata 700064, India.}
 \affiliation{Homi Bhabha National Institute, 2nd floor, BARC Training School Complex,
	Anushaktinagar, Mumbai, Maharashtra 400094, India.}
 \author{Rome Samanta}
  \email{romesamanta@gmail.com}
 % \affiliation{CEICO, Institute of Physics of the Czech Academy of Sciences, Na Slovance 1999/2, 182 21 Prague 8, Czech Republic}
% %\author{Constantinos Skordis}
% %\author{Federico R. Urban}
% \email{romesamanta@gmail.com}
 \affiliation{CEICO, Institute of Physics of the Czech Academy of Sciences, Na Slovance 1999/2, 182 21 Prague 8, Czech Republic}
 %\author{Others}

%\preprint{NUHEP-TH/19-08}

\begin{abstract} 
We show that the seesaw mechanisms that exhibit right-handed neutrino mass-dependent non-standard post-inflationary cosmology make blue-tilted inflationary gravitational waves (GW) compatible with the recent findings of nHz stochastic GW background by the pulsar-timing arrays (PTAs)  for high reheating temperatures. The right-handed neutrino (RHN) mass scale has to be $\mathcal{O}(\rm GeV)$. Remarkably, such a scenario produces a correlated signature testable by the future LIGO run. In addition to contributing to the active neutrino masses, $\mathcal{O}(\rm GeV)$ RHNs generate baryon asymmetry of the universe via low-scale-leptogenesis. They can be searched for in collider experiments. Therefore, the recent detection by PTAs is not only exciting for GWs in the nHz range; it paves the way to test and constrain well-studied mechanisms, such as seesaws, with a low-frequency and a correlated measurement of high-frequency GW spectral features, complementary to particle physics searches. 

\end{abstract}

\maketitle
%\section*{}
\section{Introduction}  Recently, pulsar-timing array (PTA) collaborations: NANOGrav, EPTA, and
PPTA, along with the InPTA, plus CPTA have released their latest data asserting significant evidence for a stochastic gravitational wave background (SGWB) at nHz frequencies \cite{NANOGrav:2023gor,Antoniadis:2023ott,Reardon:2023gzh,Xu:2023wog}. Such a finding, albeit with less statistical significance, has already been there for the last two years, creating a reasonable buzz within the scientific community \cite{ng121,ng122,ng123}. This time, however, the signal exhibits the characteristic pulsar angular correlations, known as the quadrupolar  Hellings-Downs curve \cite{Hellings:1983fr}, which is unique to an SGWB. While the sources of such GWs remain unknown, the preferred power-law $\Omega_{\rm GW}\propto f^{1.8 \pm 0.6}$, e.g., in the NANOGrav new data does not disfavor the simple GW-driven models of supermassive black hole binaries at $3\sigma$. Another exciting possibility, nonetheless, is to investigate GWs of cosmological origin. In a companion theory paper \cite{NANOGrav:2023hvm}, the NANOGrav collaboration (for definiteness, we shall focus on NANOGrav 15 yrs data \cite{NANOGrav:2023gor}; results of other PTAs are in good agreement) produced an exhaustive catalogue discussing plenty of cosmological sources that comply with the data \footnote{Unlike the 12.5 yrs NANOGrav data, which stable cosmic strings provide a good fit \cite{Blasi:2020mfx,Ellis:2020ena,rfit1,rfit2,rfit3,rfit4}, the recent data disfavors stable cosmic strings \cite{NANOGrav:2023gor}.}. Subsequently, in various articles, such sources were discussed either in the context of different cosmological models or re-performing the fit to the new data, including the results of other PTAs \cite{Ellis:2023tsl,Wang:2023len,Kitajima:2023cek,Franciolini:2023pbf,Megias:2023kiy,Fujikura:2023lkn,Han:2023olf,Zu:2023olm,Yang:2023aak,Guo:2023hyp,Shen:2023pan,Franciolini:2023wjm,Lambiase:2023pxd,Li:2023yaj,new1,new2,Athron:2023mer,Oikonomou:2023qfz,Broadhurst:2023tus}. Inflationary gravitational wave with large tensor blue-tilt (henceforth we address them as blue tilted gravitational waves--BGWs) is one of them that provides an excellent fit to the old as well as the new data \cite{ng122,NANOGrav:2023gor,bn1,bn2,bn3,bn4,Vagnozzi:2023lwo}; though, it should be noted that such BGWs can be produced in models which, in general, do not correspond to standard slow-roll inflation, see, e.g.,  \cite{bgw1,bgw2,bgw3,bgw4,bgw5,bgw6,bgw6,bgw7,bgw8}. The parameter space of such a fit is, however, restrictive. This is because GWs with large blue-tilt, considering the spectrum is still a power-law at higher frequencies, saturate Big Bang nucleosynthesis (BBN) bound on the effective number of neutrino species, disfavoring any post-inflationary cosmology founded on high reheating temperature ($T_{\rm RH}\gtrsim 10$ GeV) after inflation \cite{NANOGrav:2023hvm,Vagnozzi:2023lwo}. Nonetheless, if a non-standard matter epoch leads to entropy production between the reheating after inflation and the most recent radiation domination before the BBN \cite{Cyburt:2015mya}, BGWs get suppressed and provide a good fit to PTA data for high reheating temperatures. Now, contrary to the standard case, such a scenario allows the overall GW spectrum to span decades of frequencies with characteristic spectral features testable, e.g., by the LIGO  \cite{KAGRA:2021kbb,Peimbert:2016bdg, LIGOScientific:2016jlg}. 

In this letter, we show the seesaw mechanisms with GeV scale right-handed neutrinos (RHN), which are now being extensively discussed within the context of low-mass sterile neutrino searches (see a review: \cite{Bondarenko:2018ptm}), naturally provide an RHN mass-dependent matter epoch to fit the PTA data with BGWs (we shall see later that a high  $T_{\rm RH}$ is also a requirement of the model). The scenario correlates RHN masses with the amplitude and spectral features in the BGWs verifiable at high-frequency GW detectors, providing a novel synergic search for GeV scale RH neutrinos, which, besides generating active neutrino masses via seesaw, also offer successful baryogenesis via low-scale leptogenesis \cite{lep8,lep9,lep10,lep11}. 

The theoretical framework is founded primarily on this question: What is the origin of small (here GeV) RHN masses?  Although the electroweak naturalness condition puts an upper bound on the RHN masses: $M_i < 10^7 $ GeV \cite{nat1,nat2}, generally, in GeV scale seesaw scenarios, the origin of such small RHN masses are not addressed; they are considered to be the bare masses in the theory. Nonetheless, the seesaw Lagrangian provides all the degrees of freedom if we suppose RHN masses originate from a phase transition driven by a scalar field \cite{lsl2,Datta:2022tab}. In which case, the RHN field $N$ couples to a scalar field $\Phi$ as $\mathcal{L}\sim f_{N}N N \Phi$ (omitting family indices), with $f_N$ being a Yukawa coupling. After the phase transition, $\Phi$ obtains its vacuum expectation value $v_\Phi$ and generates RHN mass as $M = f_N v_\Phi$.  On the other hand, if kinematically allowed, $\Phi$ can decay to a pair of RHNs ($\Phi\rightarrow N N$), with the decay rate $\Gamma\propto f_N^2$. RHN masses and the decay width (or lifetime) of $\Phi$ are now connected via $f_N$. For a fixed $v_\Phi$ (which may be large), one obtains a required small value of $M$ for a small $f_N$, making $\Phi$ long-lived. We shall see that the long-lived $\Phi$ dominates the universe's energy budget as a matter component before decaying. The smaller the $M$, the longer the lifetime of $\Phi$. This results in a longer duration of matter domination--hence larger entropy production and more suppressed BGWs. The time and amount of entropy production correlate the amplitude and spectral features of BGWs with RHN mass scale $M$.  For more technical details, please see Ref.\cite{Datta:2022tab} where the idea explained above was introduced.

\section{ RHN mass-dependent matter domination}
Obtaining RHN masses via a phase transition is not ad-hoc; the coupling $f_{N}N N \Phi$ can appear as $U(1)_{B-L}$ symmetric coupling following the breaking of a grand unification group \cite{bml1,bml2,bml3,bml4}. In that case, $N$ and $\Phi$ have $B-L$ charge $1$ and $-2$, respectively. Therefore, for concreteness, we shall consider a $B-L$ phase transition, i.e., as the temperature drops, the scalar field rolls from  $\Phi=0$ towards $\Phi=v_{\Phi}$, breaking the $B-L$ symmetry. The finite temperature potential that restores the symmetry  at higher temperatures is given by \cite{Linde:1978px,Kibble:1980mv}
\bea
V(\Phi,T)=\frac{\lambda}{4}\Phi^4+D(T^2-T_0^2)\Phi^2-ET\Phi^3,\label{tmpdp}
\eea
 where 
$D,E$ and $T_0$ are functions of gauge coupling $g^\prime$, the self-interaction coupling $\lambda$, and  $v_\Phi=\frac{\mu}{\sqrt{\lambda}}$ determined from the zero temperature potential $V(\Phi,0)=-\frac{\mu^2}{2}\Phi^2+\frac{\lambda}{4}\Phi^4$ \cite{Datta:2022tab}. The last term in Eq.\eqref{tmpdp} generates a potential barrier causing a secondary minimum at $\Phi\neq0$, which at $T=T_c$ degenerates with the $\Phi=0$ one. The potential barrier vanishes at $T_0 ~(\lesssim T_c)$, making the minimum at $\Phi=0$ a maximum. The critical temperature $T_c$ and the field value $\Phi_c\equiv \Phi \left(T_c\right)$ are given by \cite{Datta:2022tab}
\bea
T_c=T_0\frac{\sqrt{\lambda D}}{\sqrt{\lambda D-E^2}},~~ \Phi_c = \sqrt{\frac{4 D}{\lambda}(T_c^2-T_0^2)}.
\eea
The transition dynamics can be described as a rolling of $\Phi$ if it smoothly transits to $\Phi=v_\Phi$, which can be quantified roughly with the order parameter $\Phi_c/T_c\ll 1$. In this case, the field rolls because the potential barrier disappears very quickly \cite{lsl2,Datta:2022tab}. The condition $\Phi_c/T_c\ll 1$ can be fulfilled for $\lambda \simeq{ g^\prime}^3$ and ${ g^\prime} \lesssim 10^{-2}$, which correspond to  $\Phi_c/T_c\lesssim 0.08$ \cite{Datta:2022tab}.  Once it rolls down, the field oscillates around $v_{\Phi}$. For  $V(\Phi)=\alpha \Phi^\beta$, the equation of state of such an oscillation phase is computed as $\omega=(\beta-2)(\beta+2)^{-1}$ \cite{Datta:2022tab}. Assuming the oscillation of the scalar field is driven by the dominant quadratic term in the potential and expanding the zero temperature potential around $v_\Phi$, we obtain $\alpha=\lambda v_{\Phi}^2$ and $\beta=2$. Therefore, the scalar field behaves like matter ($w=0$). One can also compute the angular frequency of oscillation as $m_\Phi=\sqrt{2\lambda} v_\Phi$. \\
The decay channels determine the lifetime of $\Phi$.  For $\lambda \simeq {g^\prime} ^3$ and ${g^\prime} \ll 1$, $\Phi\rightarrow z^\prime z^\prime $  is not allowed from kinematic consideration. Here $z^\prime$ is the $U(1)_{B-L}$ gauge boson. Another decay mode $\Phi\rightarrow h h$, where $h$ is the SM model Higgs, does not contribute if $\Phi$ is sequestered from SM Higgs at the tree level (even at the radiative level, it is not efficient due to the discussed small values of $f_N$). Then, the competing decay channels are $\Phi\rightarrow N N$ and $\Phi\rightarrow f\bar{f}V$ ( The corresponding two-body decay; $\Phi \rightarrow f\bar{f}$ is suppressed due to chirality flip, e.g., \cite{Han:2017yhy}), where $f$ and $V$ are SM fermions and vector bosons.  The former corresponds to a tree-level process, whereas the latter is a one-loop ($z^\prime z^\prime f$) triangle process. The strengths of these two processes are determined by the couplings $f_N$ and $g^\prime$. To control the duration of matter domination with small RHN mass via $f_N$, the process $\Phi\rightarrow N N$ should dominate ($\Gamma_{N}^\Phi\gtrsim \Gamma_{f\bar{f}V}$). In which case, the entropy produced ($\kappa$) by the decay of $\Phi$ is given by (which amounts typically equivalent to the ratio of the temperatures corresponding to the time when $\Phi$ dominates and when it decays, see, e.g., the supplementary material) \cite{Datta:2022tab}
    \bea
 \mathbb{\kappa}^{-1}\simeq \frac{\left(\frac{90}{\pi^2 g_*}\right)^{1/4} \rho_{R}\left( T_c \right)\sqrt{\Gamma_N^\Phi \tilde{M}_{Pl}}}{\rho_{\Phi}\left( T_c \right) T_c},
 \label{entrp}
 \eea
 where $\tilde{M}_{Pl}=2.4\times 10^{18}$ GeV is the reduced Planck constant, $\rho_{\Phi}\left( T_c \right) \equiv V_{\text{eff}}\left( 0,T_c\right)\simeq \frac{\lambda}{4}v_{\Phi}^4$, $g_*$ is energy degrees of freedom, and $\Gamma_N^\Phi\simeq \frac{f_N^2}{10 \pi}m_\Phi$. The analytical formula for $\kappa$ is very precise, and we shall use it to compute the GW spectrum. 
 Let’s mention the following conditions that the model complies with. I) As mentioned, $\Gamma_{N}^\Phi\gtrsim \Gamma_{f\bar{f}V}$, II) $\Phi$ decays before BBN ($T\sim 5$ MeV), III) the transition happens following reheating after inflation, i.e., $T_c\lesssim T_{\rm RH}$, IV) the vacuum energy does not dominate the radiation at $T_c$ ($\rho_{\Phi}\left( T_c \right)<\rho_R \left( T_c \right)$); violation of which leads to a second period of inflation. We shall see shortly that there are three additional constraints, excluding the PTA data, and the model has six free parameters. Therefore, the model withstands with 6:8 capacity against the constraints (including PTA data), leading to extremely robust predictions.  Specifically, we shall see that the recent PTA data can be explained only for a constrained range of RHN masses, which can be probed by the future LIGO run and is also interesting for future collider FCC-ee. 
\section{ Fit to the NANOGrav data and predictions} GWs are described with the perturbed FLRW line element:
\bea
ds^2=a(\tau)\left[-d\tau^2+(\delta_{ij}+h_{ij})dx^idx^j)\right],
\eea
where $\tau$ is the conformal time, $a(\tau)$ is the scale factor. The transverse and traceless part of the $3\times 3$ symmetric matrix $h_{ij}$; $\partial_ih^{ij}=0$ and $\delta^{ij}h_{ij}=0$, characterizes the GWs. Following the linearized evolution equation 
\bea
\partial_\mu(\sqrt{-g}\partial^\mu h_{ij})=16\pi a^2(\tau) \mathcal{\pi}_{ij},\label{lineq}
\eea
considering subdominant contribution from anisotropy stress tensor $\mathcal{\pi}_{ij}$ \cite{Zhao:2009we}, and solving the Fourier space propagation equation for $h_{ij}$, one obtains the gravitational wave energy density as \cite{WMAP:2006rnx}
 \bea
 \Omega_{\rm GW}(k)=\frac{1}{12H_0^2}\left(\frac{k}{a_0}\right)^2T_T^2(\tau_0,k)P_T(k),\label{GWeq}
 \eea
 where  $H_0\simeq 2.2 \times 10^{-4}~\rm Mpc^{-1}$ and $\tau_0=1.4\times 10^4 {\rm ~Mpc}$. The quantity $P_T(k)$ represents the primordial power spectrum connecting to the inflation models:  
 \bea
P_T(k)=r A_s(k_*)\left(\frac{k}{k_*}\right)^{n_T},
\eea
 where $r\lesssim 0.06$ \cite{BICEP2:2018kqh} (constraint V) is the tensor-to-scalar-ratio, $k=|\vec{k}|=2\pi f$ with $f$ being the frequency of the GWs at the present time at $a_0=1$, $A_s \simeq 2\times 10^{-9}$ is the scalar perturbation amplitude at the pivot scale $k_*=0.01\rm  Mpc^{-1}$ and $n_T$ is the tensor spectral index.  The simplest single-field slow-roll inflation models satisfy a consistency relation: $n_T=-r/8$ \cite{Liddle:1993fq}. We shall treat $n_T>0$ as constant, ignoring scale dependence due to higher-order corrections \cite{Kuroyanagi:2011iw}. The most important quantity in the discussion is the transfer function $T_T(\tau_0,k)$ given by \cite{t1,t2,t3,t4,t5,t6}
 \begin{equation}
\begin{split}
T_T^2(\tau_0,k)=F(k)T_1^2(\zeta_{\rm eq})T_2^2(\zeta_{\Phi})T_3^2(\zeta_{\Phi R})T_2^2(\zeta_{R}),
\end{split}
\label{fuk34}
\end{equation}
where $F(k)$ reads
\begin{equation}
\begin{split}
F(k)=\Omega_m^2\left( \frac{g_*(T_{k,\rm in})}{g_{*0}}\right)\left(\frac{g_{*s0}}{g_{*s}(T_{k,\rm in})}\right)^{4/3}\left(\frac{3j_1(k\tau_0)}{k\tau_0}\right)^2
\end{split}
\label{fuk}
\end{equation}
with $j_1(k\tau_0)$ being the spherical Bessel function, $\Omega_m=0.31$, $g_{*0}=3.36$, $g_{*0s}=3.91$. We use the scale-dependent $g_{*0(s)}(T_{k,\rm in})$  in Eq.\eqref{fuk} from \cite{gs1,gs2,t6},  where $T_{k,\rm in}$ is the temperature corresponding to the horizon entry of $k$th mode. The $T_{i}(\zeta)$s are given by 
\bea
T_1^2(\zeta)=1+1.57\zeta+ 3.42 \zeta^2,\\
T_2^2(\zeta)=\left(1-0.22\zeta^{1.5}+0.65\zeta^2 \right)^{-1},\\
T_3^2(\zeta)=1+0.59\zeta+0.65 \zeta^2,
\eea
where $\zeta_i \equiv k/k_i$, with $k_i$ s being the modes: 
\bea
k_{\rm eq}&=&7.1\times 10^{-2}\Omega_m h^2 {\rm Mpc^{-1}},
\eea
\begin{equation}
\begin{split}
k_{\Phi}=1.7\times 10^{14}\left(\frac{g_{*s}(T_\Phi)}{106.75}\right)^{1/6}\left(\frac{T_\Phi}{10^7 \rm GeV}\right){\rm Mpc^{-1}}, 
\end{split}
\label{k17r}
\end{equation}

\begin{equation}
\begin{split}
k_{\Phi R}=1.7\times 10^{14} \kappa^{2/3}\left(\frac{g_{*s}(T_\Phi)}{106.75}\right)^{1/6}\left(\frac{T_\Phi}{10^7 \rm GeV}\right){\rm Mpc^{-1}},
\end{split},
\label{k1xr}
\end{equation}
\begin{equation}
\begin{split}
k_R=1.7\times 10^{14}\kappa^{-1/3}\left(\frac{g_{*s}(T_{\rm RH})}{106.75}\right)^{1/6}\left(\frac{T_{\rm RH}}{10^7 \rm GeV}\right){\rm Mpc^{-1}}
\end{split}
\label{k1r}
\end{equation}
 crossing the horizon at standard matter-radiation equality temperature $T_{\rm eq}$, at $T_{\Phi}\simeq \left(\frac{90}{\pi^2 g_*}\right)^{1/4} \sqrt{\Gamma_N^\Phi \tilde{M}_{Pl}}$ when $\Phi$ decays, at $T_{\Phi R}$ when $\Phi$  dominates the energy density, and at $T_{\rm RH}$, respectively. Given the above set of equations and using $\kappa$ from Eq.\eqref{entrp}, we evaluate Eq.\eqref{GWeq} to obtain the GW spectrum and to fit the NANOGrav data, considering two more constraints. VI) LIGO bound on SGWB, which roughly reads $\Omega_{\rm GW} (35 \text{\:Hz})h^2\leq 6.8\times 10^{-9}$ \cite{KAGRA:2021kbb}, and VII) BBN bound on the effective number of neutrino species: $\int_{f_{\rm low}}^{f_{\rm high}}f^{-1}df \Omega_{\rm GW}(f)h^2\lesssim 5.6\times 10^{-6}\Delta N_{eff}$, where $\Delta N_{\rm eff}\lesssim 0.2$ \cite{Planck:2018vyg}. $f_{\rm low}$ corresponds to the frequency entering the horizon during BBN epoch. On the other hand, the Hubble rate at the end of inflation determines the upper limit: $f_{\rm high}=a_{\rm end} H_{\rm end}/2\pi$. For numerical computations, $f_{\rm high}\simeq 10^5$ Hz would suffice because the spectrum falls and the integration saturates.
\begin{figure*}
 	\includegraphics[scale=.38]{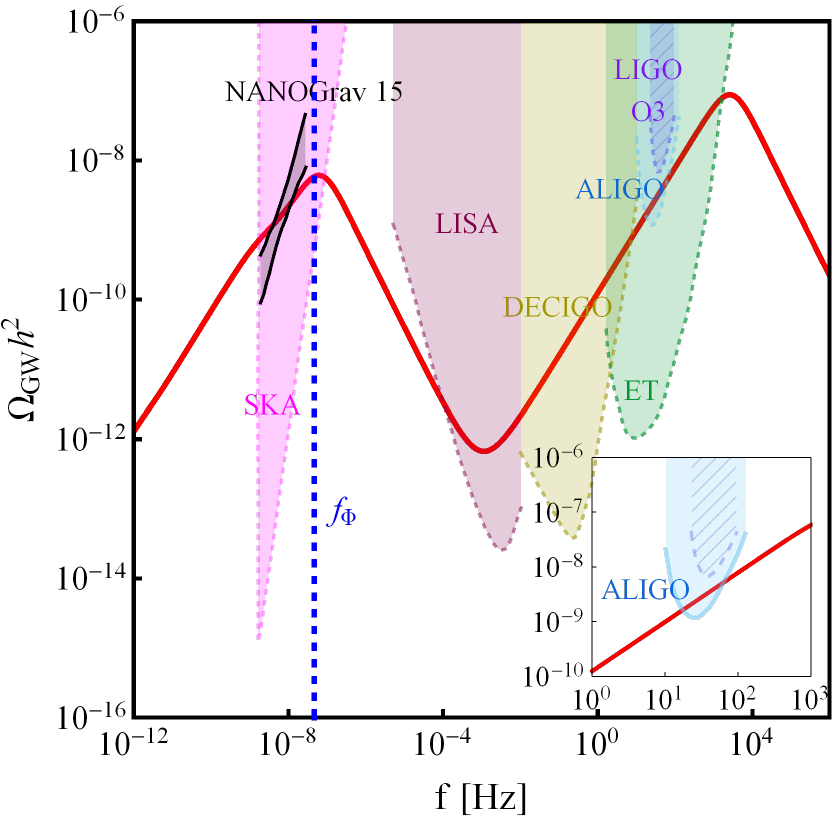}
        \includegraphics[scale=.4]{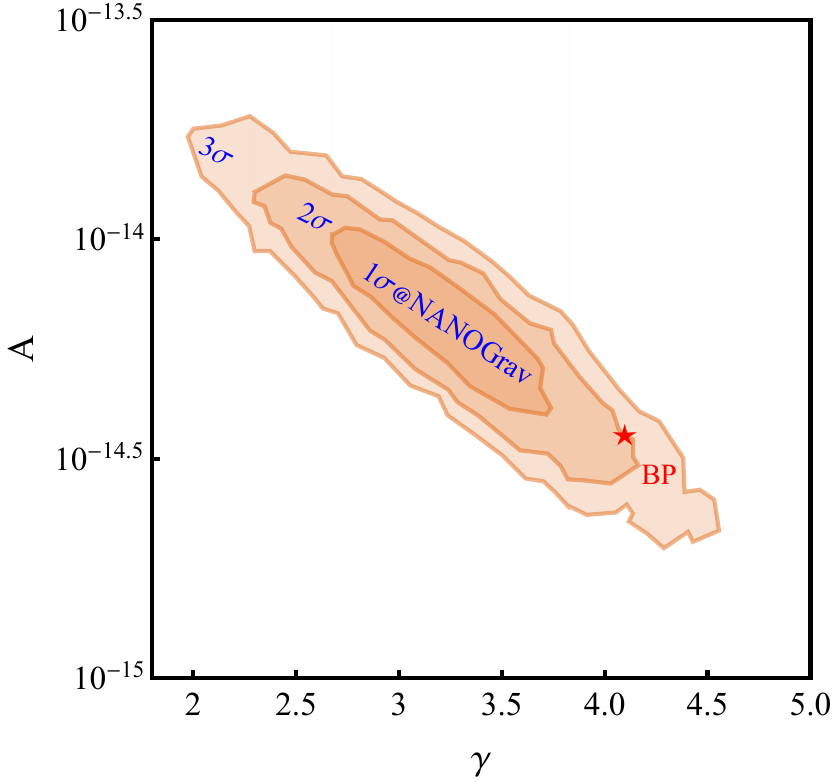} \includegraphics[scale=.45]{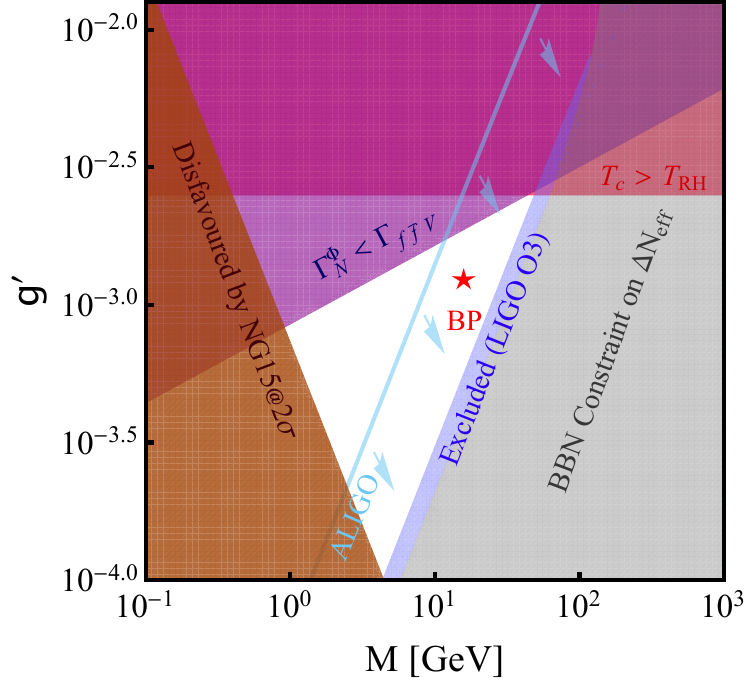}
 	\caption{ Left: The spectrum has been generated for $M=16$ GeV. For other benchmarks, please see the text. Besides NANOGrav and LIGO, sensitivities of SKA \cite{ska}, LISA \cite{lisa}, DECIGO \cite{decigo}, and ET \cite{Sathyaprakash:2012jk} are shown. The vertical dashed blue line represents the frequency corresponding to Eq.\eqref{fphi}. Middle: The red $\star$ represents the fit-point corresponding to the GW spectrum on the left. Right: An allowed parameter space on the $g^\prime-M$ plane (white). All the colored regions are excluded. The NANOGrav signal cannot be reproduced in the brown region at 2$\sigma$. BBN constraint on $\Delta N_{\rm eff}$ disfavors the grey region.  The LIGO-O3 bound on SGWB excludes the purple region. The region right to the sky-blue line (indicated by arrows) represents the future ALIGO sensitivity. In the pink region (top left corner), the three-body decay of $\Phi$ is more dominant than the RH neutrino pair production.  In the red region on the top, one has $T_c>T_{\rm RH}$.}\label{fig1}
 \end{figure*}
We follow the NANOGrav parametrization for the GW energy density to perform a power-law fit to the new data within the frequency range $f\in\left[10^{-9}~{\rm Hz},f_{yr}\right]$. The  parametrization reads
\begin{equation}
\Omega_{\rm GW}(f)=\Omega_{yr}\left(\frac{f}{f_{yr}}\right)^{(5-\gamma)}\label{pl}
\end{equation}
with $\Omega_{yr}=\frac{2\pi^2}{3H_0^2}A^2 f_{yr}^2$ and $f_{yr}=1yr^{-1}\simeq 32$ nHz. Fitting the data requires comparing Eq.\eqref{GWeq} and Eq.\eqref{pl}, then extracting the values of the amplitude $A$ and the spectral index  $\gamma$ that lie within the $1,2,3\sigma$ contours (cf. middle panel of Fig.\ref{fig1}) reported by the NANOGrav \cite{NANOGrav:2023gor}. In the left panel of Fig.\ref{fig1}, we show a GW spectrum consistent with all the constraints. To produce the figure, the following benchmark values for the model parameters have been used: $T_{\rm RH}=10^{13}$ GeV, $n_T=0.9$, $r=0.06$, $v_\Phi=10^{14}$ GeV, $g^\prime=10^{-2.9}$, and $M=16$ GeV. The corresponding values of $A$ and $\gamma$ are shown in the middle panel with the red `$\star$'. The fit is not very different from the standard BGW-fit without intermediate matter domination \cite{Vagnozzi:2023lwo}. This is because, within the NANOGrav frequency range, the transfer function $T_1 (\zeta)$ determines the spectral shape, which results in $\gamma\sim 5-n_T\simeq 4$. Note also that the first peak of the spectrum occurs at a frequency $f_\Phi>f_{yr}$ so that the NANOGrav band can be fitted with a pure power-law $\Omega_{\rm GW}(f\lesssim f_{yr}) \sim f^{n_T}$. An analytical expression for $f_\Phi$ can be derived from Eq.\eqref{k17r}, which is given by 
\bea
f_\Phi\simeq 50~ {\rm nHz}\left(\frac{M}{16~{\rm GeV}}\right)\left(\frac{10^{14}~{\rm GeV}}{v_\Phi}\right)^{1/2}\left(\frac{g^\prime}{10^{-2.9}}\right)^{3/4}.\label{fphi}
\eea

One can do a more exhaustive fit by varying the parameters. For instance, in the right panel, we varied $g^\prime$ and $M$, keeping the rest fixed to their benchmark values. Nonetheless, there is no qualitative difference  $-$ allowed values of the other parameters lie near the benchmarks. This does not change significantly even though one varies all the parameters. As stated earlier, this is because of a bunch of constraints that the model complies with. Among all the constraints, the first, i.e., $\Gamma_{N}^\Phi\gtrsim \Gamma_{f\bar{f}V}$, the third, i.e., $T_{\rm RH}\gtrsim T_c$, and the sixth, i.e., LIGO bound on SGWB, are much stronger, making the allowed parameter space consistent with the NANOGrav data, stringent. This model fits the NANOGrav data at $2\sigma$ with $M \in \left[1,47\right]$ GeV for random values of other parameters around the benchmark and without violating any constraints.
\begin{figure}
    \centering
    \includegraphics[scale=.47]{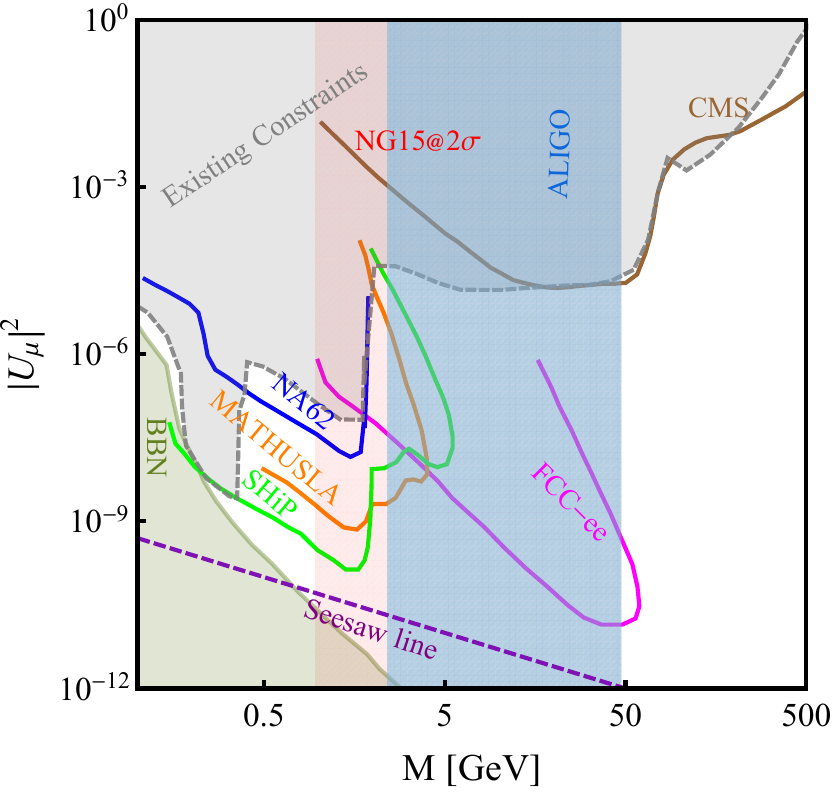}
    \caption{Particle physics exclusion and projections for RHN mixing with muon flavor. The grey region and region above the brown curve are excluded from the previous experiments and CMS 13 TeV run \cite{Bondarenko:2018ptm,Alekhin:2015byh,CMS:2018iaf}.  Future sensitivities of SHiP \cite{SHiP}, MATHUSLA \cite{MATHUSLA}, NA62 \cite{NA62}, and FCC-ee \cite{FCC} are shown with green, orange, blue, and pink curves. The region named BBN is excluded; otherwise, the decay product of RHNs would contradict BBN prediction. The dashed line represents the mixing in the canonical seesaw $|U_\mu|^2\sim m_\nu/M$.  The vertical red band represents the RHN mass range $M\sim \left[1,47\right]$ GeV consistent with NANOGrav 2$\sigma$ data. The vertical sky-blue band represents the RHN mass range $M\sim \left[2.5,47\right]$ GeV consistent with NANOGrav 2$\sigma$ data and testable with SGWB searches by advanced LIGO.}
    \label{fig2}
\end{figure}
Remarkably, consistency with the NANOGrav data makes the model extremely predictive, not only in terms of RHN masses; the infrared tail of the second peak in the GW spectrum passes through the sensitivity region of advanced LIGO (ALIGO),  for a significant portion of allowed parameter space (see the right panel in Fig.\ref{fig1}). Therefore, any non-observation of SGWB by ALIGO would potentially rule out a large parameter space of the model, {\it provided that the model fits NANOGrav data}. Not only that, the latter also motivates us to combine, for the first time, the particle physics sensitivity curves for GeV scale RHN searches with the ALIGO projection, shown in  Fig.\ref{fig2} with the vertical sky-blue band representing the RHN mass range $M \in \left[2.5,47\right]$ GeV. Particle physics experiments are sensitive to RHN masses and their mixing ($|U_{\alpha}|^2\sim m_\nu/M$) to active neutrinos, where $m_\nu$ is the active neutrino mass scale ($\simeq 0.05$ eV). On the contrary, predictions of this model depend on the former, allowing us to identify the region independent on $|U_\mu|$ in Fig.\ref{fig2}.

We conclude with the following remarks covering some additional aspects of the work.  1) The framework can be extended to other variants, such as inverse-seesaw, which offers additional phenomenology. 2)  One may straightforwardly obtain analytical expressions of the peak and dip frequencies in terms of RHN masses using Eq.\eqref{k17r}-\eqref{k1r} (via $T_\Phi$). 3) We do not present explicit computation of leptogenesis. It would be interesting to reproduce calculations, e.g., of \cite{lep11} (including entropy production), leading to a parameter space sensitive to $|U_\alpha|$ and $M$. 4) This one is perhaps the most interesting: the possibility of obtaining GWs from cosmic strings. The occurrence of $B-L$ phase transition would naturally produce cosmic gauge strings. We, however, work with unconventional values of $\lambda$ and $g^\prime$, which in general are taken $\mathcal{O}(1)$ in the Nambu-Goto simulations \cite{ngt1,ngt2,ngt3}. Additionally, for the parameter space consistent with the NANOGrav data, the cosmic string width $\delta_{w}\sim 1/\sqrt{\lambda}v_\Phi$ constitutes a considerable fraction of the horizon $H(T_c)^{-1}$ (relatively thick strings). Claiming GWs from cosmic strings in this model thus requires a straightforward assumption (which we are less confident about): results of the numerical simulations also hold for our preferred parameter range.  An existence of GWs from cosmic strings, nonetheless, would produce further spectral distortion to the BGWs shown in Fig.\ref{fig1}, making a combined peak-plateau-peak spectrum instead of a peak-dip-peak one (supplementary material can be seen). This distinguishes the model from any other matter domination+BGW scenario, even at the level of the GW spectrum.
% \begin{figure}
% %     \centering
%      \includegraphics[scale=.46]{Test_tri_NG_LSL.png}
%      \caption{ The NANOGrav data can be fitted with the region between two black mass circles (also marked with the pink band, top). The region is allowed by particle physics experiments (green band, right). LIGO-O3 bound on SGWB  rules out a portion of the region (purple band, left).  The mass range $M\sim 7-47$ GeV can probed by the ALIGO (sky-blue circle and sky-blue band, left). The red circle represents the benchmark $M=16$ GeV. The red triangle stands for the threefold testability of GeV scale seesaw models with PTAs, LIGO, and particle physics experiments.}
%      \label{fig3}
%  \end{figure}
 
\section{Summary} We discuss a novel framework to probe seesaw models with GeV scale right-handed neutrinos (RHN) with the recent pulsar timing (PTA) data interpreted as stochastic gravitational waves background (SGWB) from inflation. A fit to the PTA data with inflationary GWs predicts the mass scale of RHN to be $\mathcal{O}$(GeV) and a PTA-LIGO correlation on SGWB.  While any non-observation of SGWB by advanced LIGO (ALIGO) would rule out a large parameter space of the model, the recent PTA data motivates us to combine the particle physics sensitivity curves for low mass RHN searches with the future LIGO projection for the mass range $M\sim \left[2.5,47\right]$ GeV.  We performed the fit with the NANOGrav 15 yrs data. We do not expect a  significant qualitative change in our results if the fit is performed by combining the data from all the PTAs, because the  $A-\gamma$ global contours reported by the IPTA collaboration are similar \cite{InternationalPulsarTimingArray:2023mzf} to the NANOGrav ones.

% The findings of our work are illustrated in Fig.\ref{fig3}, broadly representing the three-fold testability
% (PTA-ALIGO-particle physics searches) of the GeV scale
% seesaw scenarios.\\

\section*{Acknowledgment}  R. Samanta is supported by the project International Mobility MSCA-IF IV FZU - CZ.02.2.69/0.0/0.0/$20\_079$/0017754 and acknowledges European Structural and Investment Fund and the Czech Ministry of Education, Youth and Sports.
%\bibliography{lesson7a1} 
%\bibliographystyle{ieeetr}

\newpage
\onecolumngrid
\begin{center}
{\bf \large Supplementary material}
\end{center}
{\bf1. Research summary for pedestrians }\\

We discuss a new method to search for GeV scale seesaw scenarios with blue-tilted gravitational waves (BGW) from inflation. This is a new indirect search for beyond the Standard Model (BSM) physics proposed in Ref.[60]  (in the letter) as a `tomographic' search. The idea is loosely inspired by computed tomography with X-rays, where in front of an X-ray source, one places an object to understand its invisible internal structures by letting the X-rays pass through the object and then studying the spectral features of the X-rays via a tomographic reconstruction. Likewise, if a propagating gravitational wave (GW) in the early universe faces any obstacle (here, a BSM theory with the potential to affect the GW propagation), then on the final GW-spectral features, properties of the obstacle (theory) get imprinted. The spectral features typically contain all the information on the timeline and time slices relevant to the BSM theory. Therefore, by studying the GW-spectral feature, one can reconstruct the phenomenon led by the BSM theory in the early universe.
\begin{figure}[H]
    \centering
    \includegraphics[scale=.65]{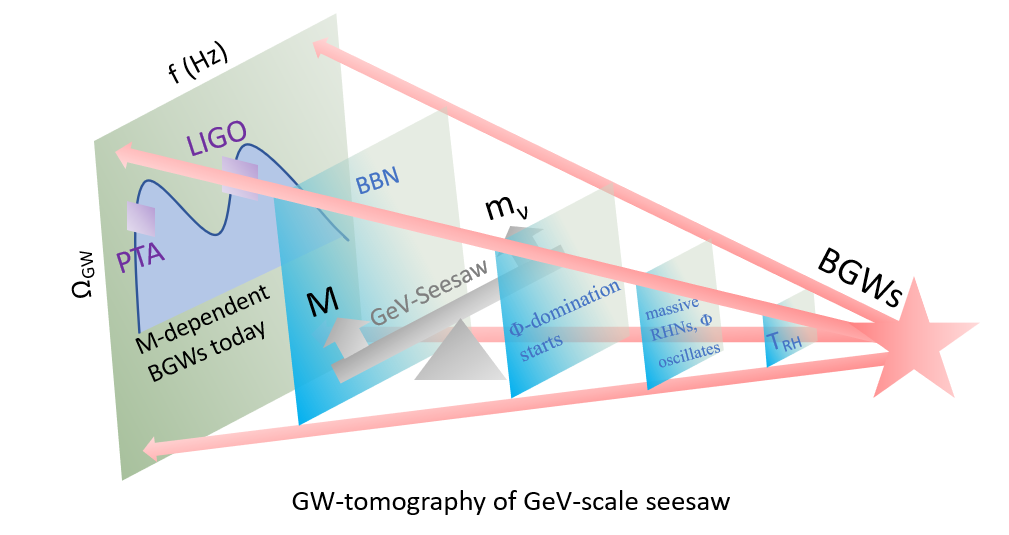}
    \caption{An illustration of the tomographic search of the seesaw mechanisms with GeV scale right-handed neutrinos. The red arrows represent the blue-tilted GWs originating from inflation. The blue squares represent different time slices in the early universe involving the GeV-scale seesaw. We expect the present-day GW spectrum with peaks and dips depending on the right-handed neutrino mass scale. Compliance with the recent PTA data implies that LIGO can test the high-frequency part of the spectrum. The present-day expectations are illustrated on the green square.   }
    \label{spfig1}
\end{figure}
In this work, the BSM theory is a seesaw mechanism of neutrino masses with GeV scale right-handed neutrinos (RHN) and blue-tilted gravitational waves from inflation do the tomography. We show that seesaws with GeV scale RHNs can create an RHN mass-dependent matter-dominated phase, affecting inflationary GW propagation. The final GW spectral feature expected at the present day is quite unique, and interestingly, the GW amplitude is large enough to explain the recent discovery of nHz stochastic gravitational wave background by pulsar timing arrays (PTAs). Also, consistency with PTA data predicts a correlated GW signal to be measured by the future LIGO runs. The RHN mass predicted in this tomographic search consistent with the PTA data is $\mathcal{O}(\rm GeV)$, which can also be tested in particle physics experiments. The overall idea is illustrated in Fig.\ref{spfig1}. \\

{\bf2. Evolution of the energy densities and entropy production}\\ 

The energy densities of radiation and the scalar field, plus the total entropy, evolve with cosmic time following the equations:
 \bea
\frac{d\rho_R}{dt}+4H\rho_R=\Gamma_N^{\Phi}\rho_{\Phi},~~
\frac{d\rho_{\Phi}}{dt}+3H\rho_{\Phi}=-\Gamma_N^{\Phi}\rho_{\Phi},~~
\frac{ds}{dt}+3Hs=\Gamma_N^{\Phi}\frac{\rho_{\Phi}}{T}.\label{be3}
\eea
We can recast them in a more numerically convenient form as
 \bea
\frac{d\rho_{R}}{dz}+\frac{4}{z}\rho_R=0, ~~
\frac{d\rho_{\Phi}}{dz}+\frac{3}{z}\frac{H}{\tilde{H}}\rho_{\Phi}+\Gamma_N^{\Phi}\frac{1}{z\tilde{H}}\rho_{\Phi}=0,\label{den2}
\eea
where $z=T_c/T$, and from the third of Eq.\eqref{be3}, the time-temperature relation has been derived as
\bea
\frac{1}{T}\frac{dT}{dt}=-\left(H+\frac{1}{3g_{*s}(T)}\frac{dg_{*s}(T)}{dt}-\Gamma_N^{\Phi}\frac{\rho_{\Phi}}{4\rho_{R}}\right)=-\tilde{H}.\label{temvar}
\eea
\begin{figure}
    \centering
    \includegraphics[scale=.6]{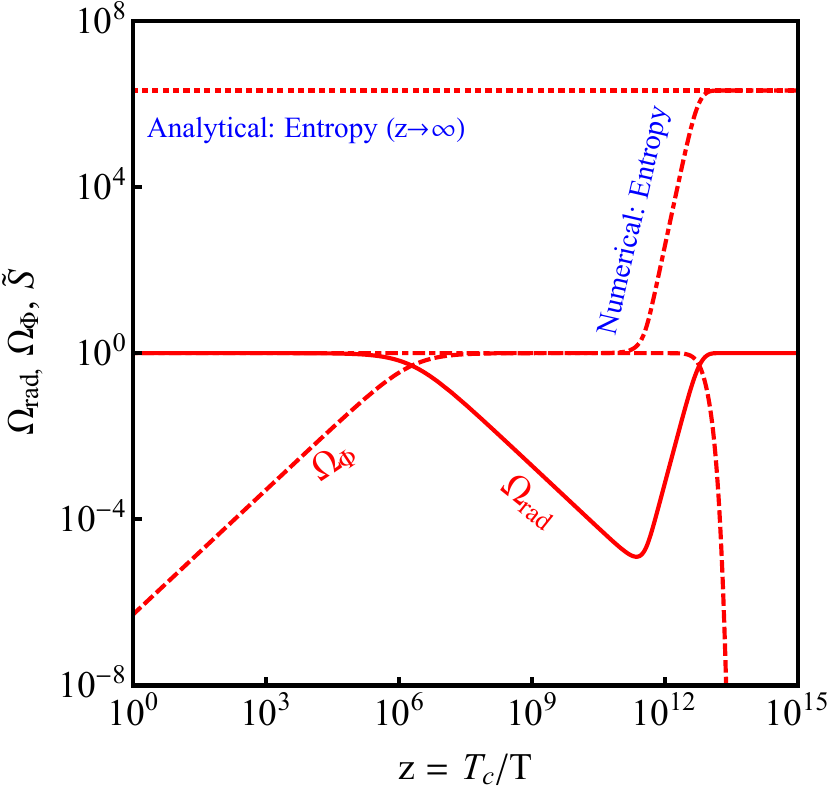}
    \caption{Evolution of the normalized energy densities of the radiation, the scalar field, and total entropy (normalized to its initial value). The horizontal dashed line represents an excellent match of the analytical approximation for the entropy production given in Eq.(3) (in the letter) with the numerical one. }
    \label{spfig2}
\end{figure}
The amount of entropy production is computed numerically by solving 
\bea
\frac{da}{dz}=\left(1+\Gamma_N^{\Phi}\frac{\rho_{\Phi}}{4\rho_{R}\tilde{H}}\right)\frac{a}{z},
\eea
and then computing the ratio of $\tilde{S}\sim a^3/z^3$ after and before the scalar field decays. The evolution of the relevant quantities with $z=T_c/T$ is presented in Fig.\ref{spfig2}.\\

{\bf3. Combined gravitational waves spectrum including cosmic strings}\\

 In this scenario, cosmic strings appear after the spontaneous breaking of  $U(1)_{B-L}$ symmetry. Following the formation, the strings get randomly distributed in space, forming close loops plus a network of horizon-size long strings. When two segments of long strings cross, they inter-commute and form loops. Long strings are characterised by a correlation length $L=\sqrt{\mu/\rho_\infty}$, with $\rho_\infty$ being the long string energy density and $\mu $ is the string tension defined as $\mu=\pi v_\Phi^2 h\left( \frac{\lambda}{2 {g^\prime}^2}\right)$. The quantity $h$ is a slowly varying function with $h\left( \frac{\lambda}{2 {g^\prime}^2}=1\right)\simeq 1$. For $\lambda\ll 2  {g^\prime}^2$, it becomes $h\left( \frac{\lambda}{2 {g^\prime}^2}\right)\simeq \left({\rm ln } ~\frac{2 {g^\prime}^2}{\lambda}\right)^{-1}$. Therefore, in our analysis, the string tension is given by $\mu=\pi v_\Phi^2 \left({\rm ln}~\frac{2}{g^\prime}\right)^{-1}$ for $\lambda={g^{\prime}}^3$. \\

Generally, a string network oscillates to enter a scaling evolution phase, characterized by stretching the correlation length due to the Hubble expansion and the fragmentation of the long strings into loops. Numerical simulations also support this phase. These loops oscillate independently and produce GWs. An attractor solution exists between these two competing dynamics: the scaling regime. In this regime, $L$ scales as cosmic time $t$, which corresponds to $\rho_\infty\propto t^{-2}$. Therefore,  the network tracks cosmological background energy density $\rho_{bg}\propto t^{-2}$ with a small constant proportional to $G\mu$, and does not dominate the energy density of the Universe.
\begin{figure}
    \centering
    \includegraphics[scale=.6]{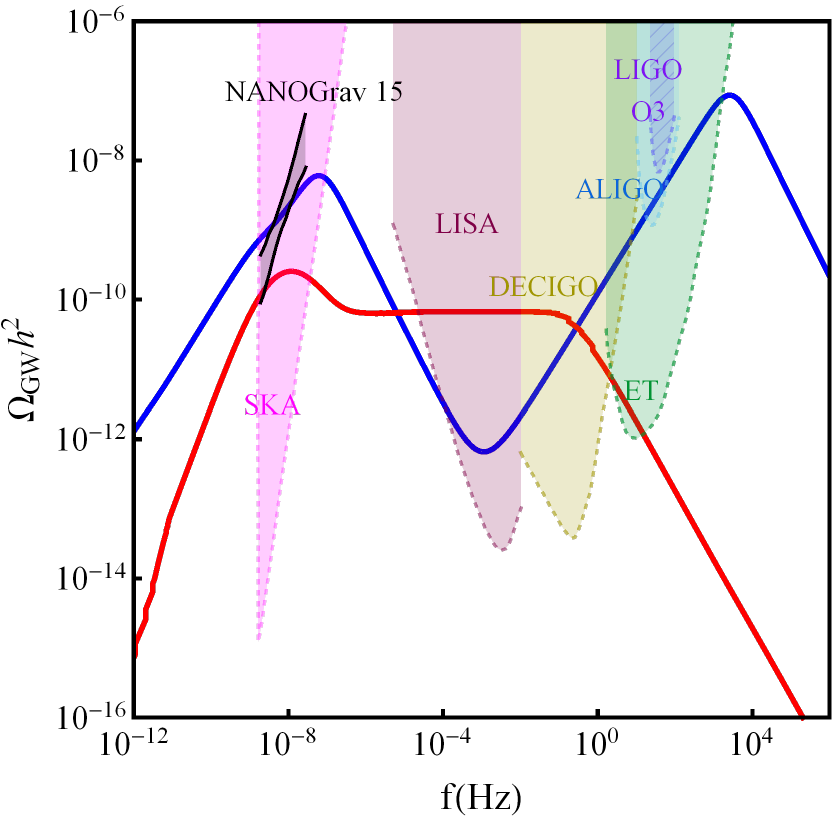}
    \caption{Blue-tilted GW spectrum (blue) combined with GWs from gauge cosmic strings (red) originating due to the $(B-L)$ symmetry breaking in this model. }
    \label{spfig3}
\end{figure}
 The evolution of a radiating loop of initial size $l_i=\alpha t_i$ is described as $l(t)=\alpha t_i-\Gamma G\mu(t-t_i)$, where $\Gamma\simeq 50$ and $\alpha\simeq 0.1$. The total energy loss from a loop can be decomposed into a set of normal-mode oscillations with frequencies $f_j=2j/l_j=a(t_0)/a(t)f$, where $j=1,2,3...j_{max}$ ($j_{max}\rightarrow\infty$). The $j$th mode GW density parameter is given by 
\bea
\Omega_{\rm GW}^{(j)}(f)=\frac{2kG\mu^2 \Gamma_j}{f\rho_c}\int_{t_{osc}}^{t_0} \left[\frac{a(t)}{a(t_0)}\right]^5 n\left(t,l_j\right)dt,\label{gwf1}
\eea
where $n\left(t,l_j\right)$ is the scaling loop number density, which can be computed analytically as well as with numerical simulations. 
Without any intermediate matter-dominated epoch, the GWs arising from Eq.\eqref{gwf1} can be described with a peak at a low frequency owing to the GW radiation from the loops  formed in the radiation era and decay in the standard matter era, plus a plateau at high frequency:
\bea
\Omega_{\rm GW}^{1,~plt}(f)=\frac{128\pi G\mu}{9\zeta(\delta)}\frac{A_r}{\epsilon_r}\Omega_r\left[(1+\epsilon_r)^{3/2}-1\right], \label{flp1}
\eea
that arises due to the loop formation and decay only in the radiation era. In Eq.\eqref{flp1}, 
$\epsilon_r=\alpha/\Gamma G\mu \gg 1$, $A_r\simeq 5.4$  and $\Omega_r\sim 9\times 10^{-5}$. \\

In the presence of a new matter epoch before the most recent radiation era, the above description remains the same; barring, the spectrum becomes red at a high frequency $f_\Phi^{\rm CS}$: $\Omega_{GW}^{1}(f>f_\Phi^{\rm CS})\propto f^{-1}$. The frequency $f_\Phi^{\rm CS}$ can be calculated analytically and is given by  
\bea
f_\Phi^{\rm CS}=\sqrt{\frac{8}{\alpha\Gamma G\mu}}t_\Phi^{-1/2}t_0^{-2/3}t_{\rm eq}^{1/6}\simeq \sqrt{\frac{8 z_{\rm eq}}{\alpha\Gamma G\mu}}\left(\frac{t_{\rm eq}}{t_\Phi}\right)^{1/2}t_0^{-1},\label{br0}
\eea
where $z_{\rm eq}\simeq 3387$ is the red-shift at the standard matter-radiation equality taking place at time $t_{\rm eq}$, and $t_\Phi<t_{\rm BBN}$ is the time when the matter domination ends.\\
In Fig.\ref{spfig3}, we show the resulting GW spectrum produced by cosmic strings ( for $j=1$ ) for the same benchmark used to produce the BGW spectrum. Notice that when combined with the BGWs, it creates a plateau in the LISA sensitivity region. This signature is exclusive to the model under consideration. The falling GW spectrum from cosmic strings can not affect the overall spectrum because, in that frequency range, the BGWs dominate. 
\end{document}